\documentclass[pra, twocolumn ,eqsecnum, superscriptaddress, amsmath, amssymb, showpacs]{revtex4}
\usepackage{graphicx}

\newcommand{\trace}[1]{\textsf{tr}(#1)}

\newcommand{\Fid}{{\overline{F}}}  
\newcommand{\braket}[2]{\left\langle #1 \vert #2\right\rangle}

\newcommand{\abs}[1]{\left\vert #1 \right\vert}
\newcommand{\Real}[1]{\textsf{Re}(#1)}

\usepackage{amsmath}
\usepackage{bbm,amsfonts}

\newcommand{\vect}[1]{\mathbf{#1}}

\def\E{{\mathcal E}}

\def\B{{\mathcal B}}

\usepackage{amsthm}
\theoremstyle{definition}

\input{Qcircuit.tex}

\begin{document}
\title{Selective and efficient quantum process tomography}

\author{Ariel Bendersky}
\affiliation{Departamento de F\'\i sica, FCEyN UBA, Pabell\'on 1, Ciudad Universitaria, 1428 Buenos Aires, Argentina}
\author{Fernando Pastawski}
\affiliation{Max-Planck-Institut f\"ur Quantenoptik, Hans-Kopfermann-Str. 1, D-85748 Garching, Germany}
\author{Juan Pablo Paz}
\affiliation{Departamento de F\'\i sica, FCEyN UBA, Pabell\'on 1, Ciudad Universitaria, 1428 Buenos Aires, Argentina}
\date{\today}

\begin{abstract}
In this paper we describe in detail and generalize a method for quantum process tomography that was presented in \cite{2008PhRvL.100s0403B}. The method enables the efficient estimation of any element of the $\chi$--matrix of a quantum process. Such elements are estimated as averages over experimental outcomes with a precision that is fixed by the number of repetitions of the experiment. Resources required to implement it scale polynomically with the number of qubits of the system. The estimation of all diagonal elements of the $\chi$--matrix can be efficiently done without any ancillary qubits. In turn, the estimation of all the off-diagonal elements requires an extra clean qubit. 
The key ideas of the method, that is based on efficient estimation by random sampling over a set of states forming a $2$--design, are described in detail. Efficient methods for preparing and detecting such states are explicitly shown.
\end{abstract}

\maketitle

\section{Introduction}

For quantum information processing to become feasible, it is necessary to be able to efficiently characterize quantum processing elements. This characterization, taking the name of {\em quantum process tomography} (QPT) is required, for instance, to design appropiate quantum error correcting codes. In general, QPT is a challenging task due to the exponential amount of parameters involved as a function of the number of qubits ($n$). Another source of complexity lies in the fact that tomographic methods are tipically indirect as the parameters characterizing a quantum proecess are not directly accessible to experiments but have to be inferred after a large number of such experiments. To be more precise, it is convenient to represent an arbitrary quantum process in an abstract way, independently of the physical carriers of information and the actual physical time required for the process to occur. This can be done by describing the process as a completely positive, linear super-operator mapping initial states into final states: $\E(\rho_{in})=\rho_{out}$. This map represents the evolution of quantum states between two snapshots in time. Trace preserving maps having the same output dimension as input dimension are of particular interest and will be the focus of our work. One possible representation of a process is through its $\chi$-matrix, which is defined with respect to a certain basis of the space of operators. Choosing the basis $\{E_m\}$ consisting of $D^2$ operators ($D=2^n$ is the dimensionality of the Hilbert space of a system of $n$ qubits), the $\chi$-matrix representation for $\E$ is such that 
\begin{equation}\label{eq:ChiRepresentation}
 \E(\rho) = \sum_{m m'} \chi_{mm'}E_m \rho E_{m'}^\dagger.  
\end{equation}
Any completely positive linear map can be written in this way. If the map preserves the trace its $\chi$--matrix is such that the condition $\sum_{m m'} \chi_{m m'} E_{m'}^\dagger E_m = I$ holds. The map $\E$ is completely characterized by the positive hermitian matrix $\chi_{m m'}$ (satisfying the above trace preserving condition).
Thus, it is clear that a complete characterization of a map $\E$  requires determining $D^4-D^2$ real parameters, a number that scales exponentially with the number of subsystems ($n$). One of the main advantages of the method we describe here is that it enables us to extract important tomographic information by investing resources that scale polynomically with $n$.

The main idea of the method, presented first in \cite{2008PhRvL.100s0403B}, is to measure any desired $\chi$--matrix element with resources scaling polynomially  with $n$. We will show that to estimate diagonal elements of  the $\chi$--matrix, the preparation and detection of states from a state $2$--design will be the main ingredient, and it will be explicitly shown. For off-diagonal coefficients, besides preparing and detecting states that go through the channel in consideration, an ancilliary qubit will be needed. The method is ``selective'' since one can use it to estimate any coefficient (or any set of coefficients). For each coefficient there is an efficient estimation strategy that we describe below. For this reason we denote our strategy as  \emph{selective efficient quantum process tomography} (SEQPT).

Our method is inspired on previous proposals that use randomized subroutines as intermediate steps for efficiently estimating any average gate fidelity. So, it is convenient to briefly describe here the features of other existing tomographic schemes. The first tomographic method proposed \cite{NielsenChuang00} is known as {Standard quantum process tomography} (SQPT). It involves preparing a set of input states $\rho_k$, and then performing full quantum state tomography on the resulting output states obtained after evolution. By doing this, we directly measure coefficients $\lambda_{jk}=Tr(\rho_k\E(\rho_j))$. However, if one wants to find the matrix elements $\chi_{mm'}$ it is necessary to invert an exponentially large system of equations relating $\lambda$ with $\chi$ \cite{chuang97prescription}. For this reason, the method is indirect (since it requires inversion to obtain matrix elements $\chi_{mn}$). It is also inefficient since, in the most general case, in order to estimate any of the coefficients $\chi_{mn}$ one needs to perform an exponentially large number of experiments and classical postprocessing. Another method known as {Direct Characterization of Quantum Dynamics} (DCQD) was recently proposed \cite{mohseni-2006-97, mohseni-2007-75} and it requires an extra ancillary system of $n$ clean qubits. The ancillary qubits must go through a clean quantum channel. Provided such a resource is available, the method enables the direct estimation of all diagonal $\chi_{mm}$ by associating them with survival probabilities of entangled (Bell) states of the system and the ancilla. The estimation of off-diagonal elements $\chi_{mm'}$ is also possible in this context but it turns out that it requires the inversion of a system of equations which, in the most general case involves an exponential number of terms. More recently, the method of {Symmetrized Characterization of Noisy Quantum Processes} \cite{JosephEmerson09282007} (SCNQP) was introduced. It is based on the idea of transforming the original channel $\E$ into a symmetrized channel $\E'$ via twirling operations. After symmetrization, only diagonal $\chi'_{mm}$ coefficients remain, being the averages over the original coefficients of the same Hamming weight. The twirling is achieved using only ($O(n)$) single qubit gates with constant depth. The values of the averaged coefficients are linearly related to output probabilities through an upper diagonal square matrix of size $n+1$. The method is ideally tailored for evaluating the applicability of relevant quantum error correcting codes \cite{2007arXiv0710.1900S} as it allows the evaluation of diagonal $\chi_{mm}$ coefficients averaged over operators of the same Hamming weight (i.e. $\chi_{00}$, average over $1$ qubit errors, etc). However, it is not possible to estimate any of the off-diagonal $\chi_{mm'}$ coefficients, which are wiped out by the symmetrization protocol, nor distinguish among specific Pauli errors of the same Hamming weight. 

Thus, existing methods do not allow the efficient estimation of an arbitrary coefficient $\chi_{mm'}$. This will be, in fact, one of the main characteristics of the method we will discuss below. Our method has a similar flavor to SCNQP adding the possibility to determine any of the coefficients $\chi_{mm'}$ (including off-diagonal ones) with polynomial resources. The method is based on {two observations}: The first, is the fact that any matrix element $\chi_{mm'}$ can be related to an average survival probability of input states under the action of the channel (or a related quantity as described below). The  average involved here is over the entire Hilbert space using the so-called Haar measure. 
The second observation is that such averages can be efficiently estimated by sampling over a finite set of states (a $2$--design, as described below). 

This paper is organized as follows. In section \ref{capSymmetrization}, we review the method we will use to compute averages over the entire Hilbert space: we define and discuss the concept of 2-designs. In section \ref{capMedicionDeCoeficientes} we present the core of the SEQPT method: We show how any element of the $\chi$--matrix can be efficiently estimated. We separately describe the estimation of diagonal and non-diagonal matrix elements presenting a detailed budget for the resources required for the estimation. 
Section \ref{capSimultanea} shows how this very same algorithm can be extended. Thus, we present a method where all the information required for the simultaneous estimation of all diagonal coefficients of the $\chi$--matrix is obtained from the same experiment. Also, we give a complete error analysis for the method.
In Appendix \ref{capMUBs} we give a brief review on mutually unbiased bases (MUBs) and show that they are a proper 2-design, and in Appendix \ref{capCambioDeBase} we show how to actually prepare any state on a complete set of MUBs (i.e. on a 2-design) by giving an explicit construction of change of basis circuits among the different bases of the MUBs. 

\section{Computing averages in Hilbert space using 2--designs}\label{capSymmetrization}

A crucial part of the method we will describe below consists in estimating averages over the entire Hilbert space of products of expectation values of two operators. The computation of quantities of this type was analized before and, as shown in \cite{renes-2004-45}, for any par of operators $O_1$ and $O_2$ we have:
\begin{equation}
\int \bra{\psi}O_1 \ket{\psi}\bra{\psi} O_2 \ket{\psi} d\psi = \frac{\trace{O_1}\trace{O_2}+\trace{O_1 O_2}}{D(D+1)}.  
\label{eq:avgastraces}
\end{equation}
The integral above is over the entire Hilbert space using the so-called Haar measure (which is the only unitarily invariant one). 

Experimentally measuring the above quantities, which involve averages over the entire Hilbert space seems completely unrealistic. However, the beautiful recent work on the theory of $2$--designs 
\cite{Dan05:MT, DCEL06:arxiv, Ambainis07, klappenecker-2005} provides the means for doing so. Delsarte \cite{Del77} showed how integrating polynomials on the sphere could be reduced to averaging the integrand on a finite set of points coined spherical designs (the important fact is that one can use the same set of points to evaluate the average of any polynomial -of a fixed degree-). The same idea can be extended to integrals over the entire Hilbert space. A state $2$--design $X$ is a set of states satisfying 
\begin{equation}
\int \bra{\psi}O_1 \ket{\psi}\bra{\psi} O_2 \ket{\psi} d\psi = 
\frac{1}{\abs{X}}\sum_{\psi \in X} \bra{\psi}O_1 \ket{\psi}\bra{\psi}  
O_2 \ket{\psi},
\label{eq:2design}
\end{equation}
for all operators $O_{1,2}$. Thus, averaging over the entire Hilbert space is equivalent to averaging over the finite set $X$ (whose cardinal is $\abs{X}$). State $2$--designs with a finite (but exponentially large) number of states exist. It is worth noticing that the computation of the exact average using a $2$--design becomes a feasible task which is still exponentially hard since the number of elements of $X$ is typically exponential in the number of qubits. However, it is now possible to realize that an estimate for the average can be efficiently found. 
Thus, this average can be estimated by randomly sampling over initial states $\ket{\psi}$ chosen from the set $X$. This idea will be crucial for the method we will present below. 

Luckily, it is rather easy to produce a state $2$--design for $n$ qubits. One possibility is to find $D+1$ mutually unbiased bases (MUB) that automatically form a state $2$--design \cite{klappenecker-2005}. Each basis will be labeled with an index $J=0,\ldots,D$ and the states within each basis will be labeled with the index $m=1,\ldots,D$. In order for the orthonormal bases to be unbiased, 
the $D(D+1)$ states of the MUBs must satisfy $\abs{\braket{\psi^J_m}{\psi^{K}_{n}}}^2 = \frac{1}{D}$ for all $J\neq K$. Since generalized Pauli operators may be partitioned into $D+1$ maximally sets of $D$ commuting operators so that each pair of sets only hold the identity $I$ as common element \cite{bandyopadhyay-new}, there are $D+1$ MUB, each one diagonalizing each of these commuting subsets of Pauli operators \cite{lawrence-2002-65}. In this way, the set of states in the MUBs can be efficiently described and also can be efficiently generated with $O(n^2)$ one and two qubit gates \cite{Ben06:TF}. It is simple to adapt the procedure used to efficiently generate any state in any MUB to compute survival probabilities of such states and also to compute the transition rates from the $(J,m)$ to $(J,m')$ states. 

It is interesting to mention that there are other sets of states that form a $2$--designs but are not MUBs. For example, Dankert et. al. \cite{Dan05:MT, DCEL06:arxiv} propose to use of approximate unitary $2$--designs (which are designs on the space of unitary operators) showing that they can be efficiently approximated. An approximate unitary $2$--design with accuracy $\epsilon + 1/D^2$ can be obtained by employing only $O(n \text{ log }\frac{1}{\epsilon})$ gates. Unitary $2$--designs acting on any fixed state induce state $2$--designs fitting into the previous scheme. Dually the action of the random unitaries may be interpreted as symmetrizing the channel $\E$ through twirling.
Following this line, we may also use weaker symmetrization protocols as in SCNQP \cite{JosephEmerson09282007} for estimating fidelities of modified channels (\ref{circ:diag},\ref{circ:offdiag}).

The importance of 2-designs for the task of quantum process parameter estimation was first pointed out by Dankert et al. \cite{DCEL06:arxiv}. This first work proved that 2-designs provide the means for efficiently measuring the fidelity of a quantum process $\Fid(\E)$ defined as 
\begin{equation}\label{eq:UnitaryAverageFidelity}
\Fid(\E) = \int_{U(D)} dU \trace{U \ket{0}\bra{0} U^\dagger \E(U\ket{0}\bra{0}U^\dagger)}
\end{equation}
where the integral is over the Haar measure.
Since the integrand is a polynomial of degree 2 in $U$ and degree 2 in $U^\dagger$
unitary 2-designs allow evaluating the expression exactly as an average over a finite set of operators $U$. If we think of the operator $U$ in the integrand as acting over $\ket{0}$, then the integral over $U$ may be cast in terms of an integral over $\ket{\psi}$.
\begin{equation}
\Fid(\E) = \int_{\psi(D)} d\psi \bra{\psi}  \E(\ket{\psi}\bra{\psi})\ket{\psi}
\end{equation}
This equation makes clear that the state $\ket{0}$ does not play a special role in defining the fidelity of the process $\E$.
At the same time, since the integrand is a polynomial of degree 2 in $\ket{\psi}$ and degree 2 in $\bra{\psi}$, it opens the possibility of using quantum state 2-designs for evaluating the average fidelity for $\E$.

\section{Selectively Measuring Channel Coefficients}\label{capMedicionDeCoeficientes}

In this section we will present the main idea that enables Selective Efficient Quantum Process Tomography (SEQPT). Below, we will separately discuss the evaluation of diagonal and non-diagonal elements of the $\chi$ matrix. However, the evaluation of both type of coefficients is based on the use of a mathematical identity that relates such coefficients with an average fidelity of a modified channel. Thus, using equation \eqref{eq:avgastraces} above, together with the $\chi$ matrix representation of the channel $\E$, it is simple to show the validity of the following equation:
\begin{equation}
\label{eqFidel}
 F_{ab}\left(\E \right)=\int_{\psi(D)} \bra{\psi}E_a\E\left(\ket{\psi}\bra{\psi} \right)E_b \ket{\psi}d\psi = \frac{D\chi_{ab}+\delta_{ab}}{D+1}.
\end{equation}
This equation is valid provided we use an operator basis $\left\lbrace E_m \right\rbrace$ which is orthogonal ($\trace{E_m E_n^{\dagger}} = D \delta_{mn}$) and such that $\trace{E_m} = D  \delta_{m0}$. An example of such kind of basis is the one formed by the generalized Pauli operators (obtained by $n$--fold tensor products of the identity and/or the three Pauli operators on each qubit). The above equation is easy to prove and is the key of our method. 

\subsection{Evaluating diagonal coefficients}

\begin{figure}[h]
\begin{equation*}
\Qcircuit @C=1em @R=.7em {
     \ket{\psi} & &  {/} \qw &\gate{\mbox{\huge $\E$}} & \gate{E_m^\dagger} &\qw & \meter   
}
\ket{\psi}\bra{\psi} 
\end{equation*}
\caption{Circuit for measuring $\chi_{mm}$ for a given channel $\E$.}
\label{circ:diag}
\end{figure}
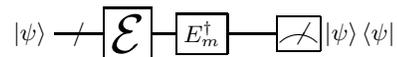

In particular, the above equation shows that diagonal coefficients $\chi_{mm}$ are directly related to averaged fidelities of the modified channel $\E_m$ defined as  $\E_m\left(\rho\right)=E_m^{\dagger}\E\left(\rho\right)E_m$. Thus, it is straightforward to show that
\begin{equation}
 F_{mm}\left(\E_m \right)=\frac{D\chi_{mm}+1}{D+1}.
\end{equation}
The channel $\E_m$ is simply obtained as the application of the original channel $\E$ followed by the operator $E_m^{\dagger}$ (which we assume to be unitary in what follows). Hence, if a method for measuring fidelity is available, diagonal elements of the $\chi$--matrix are also accessible. One such method is shown in figure \ref{circ:diag}.
According to our previous discussion, the fidelity averaged over the entire Hilbert space can be evaluated also by averaging over a state 2-design. Thus, the diagonal coefficients are evaluated as the average fidelity over all the states of the 2-design. Below, we will show that estimating the average fidelity with fixed precision requires a number of measurements that scales polynomially with the number of qubits. To do this, one should simply be able to randomly sample over the states of the 2-design. In such case the standard deviation in the estimation decreases as $1/\sqrt{M}$ where $M$ is the number of experimental runs. Also, in the Appendix \ref{capMUBs}  we explicitly show an efficient method to prepare the set of states of a 2-design. This completes the method to evaluate diagonal coefficients. 

\subsection{Evaluating off-diagonal coefficients}

The evaluation of off-diagonal elements requires a slightly different strategy. This is the case because, unlike diagonal coefficients, $\chi_{mm'}$ is not related with the average fidelity of a physically realizable (completely positive) channel. In fact, for $m\neq m'$ the above expression reduces to    
\begin{equation}
 \int \bra{\psi}\E\left(E_m^{\dagger}\ket{\psi}\bra{\psi} E_{m'} \right)\ket{\psi}d\psi = \frac{D\chi_{mm'}}{D+1}.
\end{equation}
As $\rho\rightarrow \E\left(E_m^{\dagger} \rho E_{m'} \right)$ is in general not a physical map (it is not completely positive nor trace preserving) the measurement of off-diagonal coefficients is not as straightforward. However, as we will now show, it can be achieved by using a single qubit as an ancilla. 

The measurement scheme is very similar to the one used in the DQC1 model of quantum computation \cite{2002Natur.418...59M} and is described by the circuit exhibited in Figure \ref{circ:offdiag}. As before, the state $\ket{\psi}$ used as input should be randomly chosen from a 2-design. 
\begin{figure}[ht]
\begin{equation*}
\Qcircuit @C=.9em @R=.4em {
     \ket{0}_{\text{Ancilla}}&& & & \gate{H} &
         \ctrl{1} & \ctrlo{1} &
        \qw & \meter & \sigma_x
     \\
     \ket{\psi}_{\text{Main}}&& & & {/}\qw &   
         \gate{E_m^\dagger} & \gate{E_{m'}^\dagger} & \gate{\mbox{\huge $\E$}} &
          \meter & ~\,\,\,\ket{\psi}\bra{\psi}
\gategroup{1}{5}{2}{8}{.7em}{--}
}
\end{equation*}
\caption{Circuit for measuring $\Real{\chi_{mn}}$ for a given channel $\E$}
\label{circ:offdiag} 
\end{figure}
Let us now analyze the circuit \ref{circ:offdiag} to show that it indeed measures $\chi_{m m'}$. 

The input state is $\rho_0=\ket{0}\bra{0}\otimes \ket{\psi}\bra{\psi}$.
After going through the circuit, but prior to measurement, the state is given by
\begin{eqnarray}
 \rho_f =&&  \frac{1}{2}\left( \ket{0}\bra{0}\otimes E_{m'}^{\dagger} \ket{\psi}\bra{\psi} E_{m'}+\right.\nonumber \\
&+& \ket{0}\bra{1}\otimes E_{m'}^{\dagger} \ket{\psi}\bra{\psi} E_m + \nonumber \\
&+&   \ket{1}\bra{0}\otimes E_{m}^{\dagger}  \ket{\psi}\bra{\psi} E_{m'}+ \nonumber \\
&+&\left. \ket{1}\bra{1}\otimes E_{m}^{\dagger}  \ket{\psi}\bra{\psi} E_m \right).
\end{eqnarray}
It can be easily shown that the expectations value of $\sigma_x$ or $\sigma_y$ on the ancilla qubit conditioned to the survival of the state  $\ket{\psi}$ on the main system is related to the off-diagonal $\chi_{mm'}$ coefficients as
\begin{eqnarray}
 \int  \trace{\rho_f\left(\sigma_x\otimes \ket{\psi}\bra{\psi} \right)} d\psi &=& \frac{D\text{Re}\left(\chi_{mm'} \right) +\delta_{mm'}}{D+1}\\
\int  \trace{\rho_f\left(\sigma_y\otimes\ \ket{\psi}\bra{\psi} \right)} d\psi &=& \frac{D\text{Im}\left(\chi_{mm'} \right)}{D+1} .
\end{eqnarray}
This shows how to measure off-diagonal $\chi$ coefficients.

\section{Error analysis and Generalizations}\label{capSimultanea} 

\subsection{Error analysis for selective and efficient quantum process tomography}

\begin{figure}[ht]
\begin{equation*}
\Qcircuit @C=1em @R=.7em {
     \Pi_{J,\vec{k}} & &  {/} \qw &\gate{\mbox{\huge $\E$}} & \gate{E_{m_0}^\dagger} &\qw & \meter   
}
\Pi_{J,\vec{k}}
\end{equation*}
\caption{2-design circuit for measuring $\chi_{mm}$ for a given channel $\E$}
\label{circ:diag2}
\end{figure}
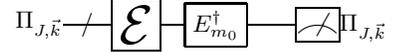

The method discussed above requires the use of states forming a 2-design as inputs and the detection of such states to estimate the survival probabilities. The use of 2-designs is crucial to evaluate the required average. In particular, it is useful to use a special type of 2-design formed by the $D(D+1)$ states belonging to $D+1$ mutually unbiased bases of the Hilbert space. We will denote projector onto the $\vec{k}$-th state of the $J$'th base as $\Pi_{J, \vec{k}}$ (see Appendix \ref{capMUBs}). Using such notation, the circuit for measuring the circuit in  figure \ref{circ:diag2} describes the protocol for the estimation of the $\chi_{m, m}$, which is obtained by randomly sampling over $J$ and $\vec{k}$. 
An effective way of estimating this average, known as Monte Carlo sampling, is to randomly choose both $J$ and $\vec{k}$ for each experiment and take an average over a set of $M$ such experiments.
The values obtained in this way are unbiassed estimators of $\chi_{m_0 m_0}$.
Furthermore, since the results of each individual experiment are either $0$ or $1$, we can bound the variance of the average over $M$ experiments by $\frac{1}{4M}$. If we want to ensure with a probability $p$ that the error is lower than $\epsilon$, then a Chernoff bound implies that a number of experiments satisfying $M\geq \text{ln}[2( 1-p )^{-1} ] / (2\epsilon^2)$ is needed. Each of these $M$ experiments will have a complexity of $O(n^2)$ arising from the number of elementary quantum gates required for the change of basis circuit (See Apendix \ref{capCambioDeBase}). The estimation of off-diagonal elements of the $\chi$--matrix requires four times more experiments because of the measurement on the ancilliary qubit.

\subsection{Generalization I: Simultaneous estimation of diagonal coefficient using transition probabilities}

It is useful to notice that the method can be easily extended if one is able to prepare the states corresponding to the MUBs associated with the operator basis $E_m$ used to describe the channel. Let us focus on the case where the operators $E_m$ are generalized Pauli operators. Suppose that the projectors $\Pi_{J, \vec{k}}$ are the states stabilized by commuting subgroups of Pauli operators (see Appendix \ref{capMUBs}). Then, we can simply find out how the operator $E_{m_0}$ acts on the state $\Pi_{J, \vec{k}}$ upon conjugation. Hence, the expectation value of the observable:
\[
\trace{E_{m_0}^\dagger \E(\Pi_{J, \vec{k}}) E_{m_0} \Pi_{J,\vec{k}}}
\]
corresponding to instances of the experiment is equal to:
\begin{equation}\label{diag_trans}
\trace{\E(\Pi_{J, \vec{k}}) \Pi_{J,\vec{k'}}}
\end{equation}
for some $\vec{k'}$ which depends on $\vec{k}$, $J$ and $E_{m_0}$. 
Then, it is clear that by detecting not only the survival of the state but also recording all the possible transitions we can obtain all the required information to estimate any diagonal coefficient and not a single one. Thus, the strategy is simple: one maintains the random preparation step of a state labeled by the indices $(J,\vec{k})$ and one stores the information about the recorded state labeled with $(J,\vec{k'})$. Given the operator $E_m$, the event should be counted in the estimation of $\chi_{mm}$ only if the input state $(J,\vec{k})$ is mapped onto the output state $(J,\vec{k'})$ by the action of $E_m$.  
Suppose that we perform a set of repetitions of this experiment and denote each event by a triple $(J, \vec{k}, \vec{k'})$, where $J$ indicates the basis randomly chosen for the experiment, $\vec{k}$ the randomly chosen input state and $\vec{k'}$, the measured output state. Then, the experiment should be counted positively in the fidelity of $E_m$ if and only if $\vec{k'}-\vec{k}$ is precisely the commutation vector of the operator $E_m$ with respect to basis $\B_J$. The commutation vector of an operator $E$ respect to basis $\B_J$ is the binary vector $\vec{v}$  such that 
\begin{equation}\label{eq:comutation_vector}
 J_i E  = (-1)^{v_i} E J_i ,
\end{equation}
where $J_1, J_2, \ldots, J_n$ are the canonical generators for the stabilizer group of $\B_J$.
We may restate this as saying that $v_i$ is the simplectic inner product between the vector describing $E$ and the one describing $J_i$.

Since the commutation vector is so important for the estimation of the $\chi_{mm}$ coefficients, we will show how it can be efficiently calculated from the canonical descriptions of $E_m$ and $\B_J$.
For $E_m$, we assume a canonical description through a binary vector with two parts, $\vec{m_X}$ and $\vec{m_Z}$ such that $E_m = X^{\vec{m_X}}  Z^{\vec{m_Z}}$.
For $\B_J$, there are two possibilities.
Either $\B_J$ is the computational basis $\B_Z$ or $\B_J$ is a basis stabilized by the group described by the binary vector $\vec{J}$ as in equation \eqref{ecEstab} from Appendix \ref{capMUBs}.
In either case, a canonical representation of the $n$ generators of the stabilizer may be obtained with only $O(n^2)$ operations.
The calculation of the commutation vector then additionally requires the calculation of $n$ simplectic inner products thus maintaining the algorithmic complexity.

Thus, the estimation of a specific $\chi_{mm}$ from such a set of $M$ experiments requires $O(Mn)$ storage for the description of the $M$ individual experiments.
The amount of classical processing required is $O(M n^2)$, mainly for the verification of $M$ commutation vectors.
Since each of the $M $ experiment already requires $O(n^2)$ elementary quantum gates and $O(n^3)$ classical processing to determine these quantum gates, this will be the dominant complexity term for the estimation of $\chi_{mm}$.

\subsection{Generalization II: Detecting and measuring large $\chi_{mm}$ coefficients}

We have shown that a single set of experiments is capable of providing information to estimate any of the diagonal $\chi_{mm}$ coefficients.
We will now extend this result to determine the operators $E_m$ related to the largest $\chi_{mm}$ coefficients.
A straightforward search by estimating all $\chi_{mm}$ coefficient is not a reasonable approach to doing this, as the number of such coefficients is exponential in $n$.
Surprisingly, we will see that finding and estimating the set of such $\chi_{mm}$ is actually possible under the condition that there be only \emph{a few} $\chi_{mm}$ with high values.
This is precisely the case that can be effectively remedied by quantum error correction, where the set of correctable error syndromes is typically small.

Suppose that we estimate all 
$F\left(\E_m \right)=\frac{D\chi_{mm}+1}{D+1}$ coefficients from a set of $M$ experiments.
The smallest nonzero value we could obtain in such an estimation is $\frac{1}{M}$, while the second smallest value is $\frac{2}{M}$ and so on.
The efficiency of our criteria for finding the $E_m$ with large $\chi_{mm}$ coefficients is based on the fact that we can efficiently detect all $F\left(\E_m \right)$ for which the estimated value is greater or equal to $\frac{2}{M}$.
Furthermore, we will be able to quantify how unlikely it is for a $F\left(\E_m \right)$ larger than $\epsilon$, to have an estimated value smaller than $\frac{2}{M}$.

Suppose we wish to determine the operator $E_m$ consistent with two experiments $(J, \vec{k_1}, \vec{k_2})$, $(J', \vec{k'_1}, \vec{k'_2})$.
If $J = J'$ there will be either $0$ or $D$ such operators $E_m$.
Otherwise, there is exactly one such operator which we will show how to determine.
We will start by noting that any operator $E_m$ can, up to a phase, be written as:
\begin{equation}\label{operator_refactoring}
 E_m \cong \prod_{i=0}^{n-1} J^{q_i}_i \times \prod_{i=0}^{n-1} J'^{q'_i}_i
\end{equation}
Where the $J_i$ and $J'_i$, with $i \in \{1\ldots n\}$, are the canonical generators of the stabilizer $J$ and $J'$ respectively.
Once the vectors $\vec{q}$ and $\vec{q'}$ are obtained, it is straightforward to obtain the canonical representation for $E_m$ with $O(n^2)$ classical operations.
Note that representation (\ref{operator_refactoring}) of $E_m$ is just a generalization of the canonical representation, where $\B_J=\B_X=\B_\vect{0}$ and $\B_{J'}= \B_Z$.

We must obtain $\vect{q}$ and $\vect{q'}$ such the $E_m$ given by representation \ref{operator_refactoring} satisfies:
\begin{equation}\label{eq:condicionDobleError}
\begin{array}{rcl} 
J_i E_m &=& (-1)^{k_2-k_1} E_m J_i \\
{J'}_i E_m &=& (-1)^{k'_2 - k'_1} E_m {J'}_i
\end{array}
\end{equation}
To do this, we determine the non-singular binary matrix $C$ such that:
\begin{equation}
 J_i {J'}_j = (-1)^{C_{i,j}} {J'}_j J_i
\end{equation}
Condition \ref{eq:condicionDobleError} may now be translated as:
\begin{equation}
\begin{array}{rcl}
 \vect{k_2}-\vect{k_1} &=& C \vect{q'}\\
 \vect{k'_2}-\vect{k'_1}&=& C^T \vect{q}
\end{array}
\end{equation}
Thus, by inverting $C$ and $C^T$, we may obtain the necessary values for $\vect{q}$ and $\vect{q'}$.
This procedure may be repeated on each of the $\binom{M}{2}$ pairs of experiment data to find the $E_m$ for which the estimate for $F(\E_m)\geq 2/M$.

\subsection{Error analysis for simulaneous determination of coefficients}

One may consider now how many experiments are needed to obtain the $\chi_{mm}$ coefficients with a certain precision.
If one wishes to measure all $\chi_{mm}$ coefficients greater than $\epsilon$, all within individual uncertainty $\delta$, one may give the number of experiments $M$ sufficient for achieving this with a probability $P$ as:
\begin{equation}\label{eq:CondicionMdelta}
 M \geq \frac{2\left(D+\frac{1}{\epsilon}\right)(D+1)}{D^2 {\delta}^2 (1-P)} = 
\frac{2\left(1+\frac{1}{D\epsilon}\right)(1+\frac{1}{D})}{{\delta}^2 (1-P)}
\end{equation}
If one further considers $\epsilon \gg \frac{1}{D}$ this expression may be simplified to:
\begin{equation}\label{eq:CondicionMdeltasimplificada}
 M \gtrsim \frac{2}{\delta^2(1-P)}
\end{equation}

This means that we may perform full diagonal tomography with only polynomial resources in both the number of qubits in the system and the desired precision $\delta$.
A loophole in this argument is that for a random channel the coefficients $\chi_{m m}$ will actually be expected to take typical values close to $\frac{1}{D}$.
The proposed method will yield good results when the channel under consideration is not random, particularly for characterizing highly local noise.

\section{Summary}

In this article we have shown how to selectively and efficiently measure any coefficient of the $\chi$--matrix representation of a channel. To estimate such coefficients we could adapt any method capable of efficiently estimate the average fidelity of a channel. In particular, we described in detail how to do this by estimating the fidelity of states randomly sampled over the states of a set of mutually unbiased basis, which have the property of forming a state 2-design.
We presented an explicit construction of the change of basis circuits for such a MUB set composed by bases stabilized by tensor product Pauli operators.
The fact that these MUBs form a 2-designs allows us to sample over a finite set of states to obtain the mean values required for the coefficient estimation.
We are also able to profit from the rich stabilizer properties of this construction to allow the estimation of any diagonal $\chi_{mm}$ from the same set of experimental measurements.
This construction, allowing the efficient and selective estimation of coefficients is not the only strength of this method. 
It also enables us to efficiently determine every diagonal coefficient larger than a certain value.
To our knowledge, this is the first application profiting from both the 2-design averaging property and stabilizer properties of this set of states.

\appendix

\section{Mutually Unbiased Bases}
\label{capMUBs}

Mutually unbiased bases (MUB) are a construct from combinatorial mathematics that has become part of the theory toolbox for quantum information.
We say that two orthonormal bases $\B_J=\{\ket{\psi^{J}_m}:m\in 1..D\}$ and $\B_K=\{\ket{\psi^{K}_l}:l\in 1..D\}$ are mutually unbiased iff
\begin{equation}
 \abs{ \braket{\psi^{J}_m}{\psi^{K}_l}   }=\frac{1}{D}.\,\,\,  \forall m, l
\end{equation}
The usual reading of this equation states that measurement in basis $\B_J$ gives absolutely no information about measurement in basis $\B_K$ and vice versa.
It has been shown that there can be at most $D+1$ bases which are mutually unbiased and constructions are only known for prime power dimensions.
Klappenecker and Roetteler \cite{klappenecker-2005} related maximal sets of MUB to 2-designs.
They proved that the set of states belonging to $D+1$ mutually unbiassed bases is itself a state 2-design.
Earlier, Bandyopadhyay et al.\cite{bandyopadhyay-new} had proven a strong connection between sets of mutually unbiased bases and maximally commuting sets of orthogonal unitary operators.
One of their results is that if one partitions a complete set of $D^2-1$ mutually orthogonal traceless operators into $D+1$ subsets of $D-1$ commuting operators each, then the $D+1$ bases diagonalizing each of these subsets are mutually unbiased.
We will provide an explicit construction of such sets as it will later be necessary to refer to it and invoke some of its properties.

When dealing with tensor product Pauli operators, the construction going from operators to states and back may be cast in terms of the stabilizer formalism \cite{1997PhDT........32G, 1998quant.ph..7006G, 2000quant.ph..4072G}.
In this formalism we will say that a state $\ket{\psi}$ is stabilized by an operator $O$ if it holds that $O\ket{\psi}=\pm\ket{\psi}$. States are described by the set of operators stabilizing them and the corresponding eigenvalues.
If the $2^{2n}-1$ generalized Pauli operators are partitioned into $2^n+1$ maximally conmuting subsets of $2^n-1$ Pauli operators, each of these subsets will be the stabilizer of a basis, and each of these bases will be unbiased to each other. Thus, the problem of finding the stabilizer groups for the MUB set is reduced to that of partitioning the generalized Pauli operators into $2^n+1$ Abelian groups.

The easiest way to construct this partition is via the finite field construction first introduced by Wootters \cite{wootters} and used by Paz et al. \cite{2005PhRvA..72a2309P}. The first requirement is the construction of the companion matrix $M$ for a primitive polynomial of the finite field $GF(2^n)$:
\begin{equation}
 M=\left( \begin{array}{cccccc}
           0&1&0&0&\cdots &0 \\
	   0&0&1&0&\cdots &0 \\
	   \vdots& & &\ddots & &\vdots \\ 
	   0& & & & & 1     \\
           r_0&r_1&r_2& \cdots & & r_{n-1}        \end{array}\right)
\end{equation}
where the primitive polynomial for the finite field is $p\left(x\right)=r_0 + r_1 x +r_2 x^2+...+r_{n-1}x^{n-1}+x^n$. 

This matrix has the property that $M^D=M$ and $M^k\neq M$ $\forall k<D$, where every operation on the matrix is performed modulo 2.
Consider the following sets of generalized Pauli operators,
\begin{equation}\label{ecEstab}
G_{\vec{b}}=\left\lbrace 1, P_{\vec{b},j}=X^{\vec{1}M^j}Z^{\vec{b}\hat{M}^{j}}: j= 1,..., D-1  \right\rbrace,
\end{equation}
where $\vec{b} \in \{0,1\}^n$ is an $n$ bit vector, $\hat{M}$ is the transpose of $M$, $\vec{1}=\left(1, 0, 0, ... \right)$ is the first canonical binary basis vector and $X^{\vec{b}}=\bigotimes_i X^{b_i}$.
Note that since $M^D=M$, we have that $j=0$ is equivalent to $j=D-1$.

It is easy to check that $G_{\vec{b}}$ is an Abelian group and that the only common operator between groups $G_{\vec{b}}$ and $G_{\vec{b'}}$ is the identity. Thus, the sets $G_{\vec{b}}$, along with the group formed by the 
tensor product of $Z$ operators -that is, the stabilizer group for the computational basis- is the partition needed.

This completes the explicit construction of the MUB set, and of the required 2-design. On Appendix \ref{capCambioDeBase} we will see an operational approach on how to prepare the states from this MUB set, and how to measure on each of these bases.

\section{Change of basis circuit}\label{capCambioDeBase}

The circuits presented in section \ref{capMedicionDeCoeficientes} assume that states are sampled over the whole Hilbert space. 
However, it is sufficient to sample over a complete set of MUBs, since these form a state 2-design.
In this appendix we will show how to build efficient change of basis circuits for a given set of MUBs that, along with translations in the computational basis, will allow to sample over every state from the 2--design.
The change of basis circuits should implement the unitary rotation $V_J^K$ such that
\begin{equation}
V_J^K \ket{\psi_a^J}=\ket{\psi_a^K}, \forall a .
\end{equation}

These circuits are a main component of the tomographic scheme introduced in this work, since they are used for the preparation of arbitrary MUB states and to measure in non-computational MUB basis.
Hence, the efficiency achieved in this step will be reflected in that of the whole method.

The construction is divided into two stages:

\begin{enumerate}
 \item The construction, for a $J$ basis, of the circuits $T_a^b$ such that $T_a^b \ket{\psi_b^J}=\ket{\psi_a^J}$. This construction is trivial if the chosen $J$ corresponds to the computational basis.
 \item The construction of the $V_J^K$ for the same fixed $J$, since every other change of basis can be built by combining two of these circuits via $V_L^M=V_J^M V_L^J={V_M^J}^{\dagger} V_L^J$.
\end{enumerate}

With an efficient solution to the second stage, it will be possible to efficiently go from any state of any of the basis from the MUB set to any other state of any other basis from the set. Thus solving the problem of averaging over a 2--design.

\subsection{Circuits for the change of basis}

In this section we will present an efficient construction for circuits  that convert any state from the computational basis into the corresponding state of the basis stabilized by $G_{\vec{b}}$.

The problem of finding a change of basis quantum circuit to go from the computational basis to any other is equivalent to that of finding a circuit that, under conjugation, transforms any tensor product of local $Z$ operators -that is, the stabilizers of the computational basis- into the corresponding stabilizers of the target basis (i.e.: the operators belonging to $G_{\vec{b}}$). 
In fact, we are looking for a unitary operator $V_{\vec{b}}$ such that:

\begin{equation}
 P_{\vec{b},k} V_{\vec{b}} \ket{i}=\pm V_{\vec{b}} \ket{i}, \forall i, \forall k
\end{equation}
where $P_{\vec{b},k}\in G_{\vec{b}}$ and $\ket{i}$ is the state $i$ from the computational basis. So $V_{\vec{b}}^{\dagger}P_{\vec{b},k} V_{\vec{b}}$ has to be, for every $k$, a stabilizer operator for the computational basis.

The first step for the construction is to find a set of generators for the stabilizer group of the target basis. This can be done easily with the definition (\ref{ecEstab}) using $j=0, ..., n-1$, and it is efficient since it only requires $O(n^2)$ classical operations.
This is thanks to the fact that $M$ is sparse allowing multiplication of vectors by $M$ to be performed with only $O(n)$ operations.
Then, in the following steps, each of the operators is going to be conjugated into stabilizers of the computational basis.

The second step is to chose the first operator from the generator group. That is, given the generator group in the form of equation (\ref{ecEstab}), take the operator $P_{\vec{b},0}$, where $\vec{b}$ labels the generator group under consideration.

As the third step for the construction, single qubit rotations should be performed on each qubit to transform the operator into a product of single qubit $Z$ and $1$, as follows:

For each qubit:
\begin{itemize}
 \item If the operator has a $1$ on the qubit in question, nothing should be done.
 \item If the operator has a $Z$, nothing should be done.
 \item If the operator has an $X$, Hadamard conjugation should be performed.
 \item If the operator has a $Y$, phase (for a single qubit in the computational basis it acts as $T\ket{b}=i^b\ket{b}$) and Hadamard conjugation should be performed.
\end{itemize}

This step is summarized as follows: the operator chosen should be conjugated by

\begin{equation}
 S=\bigotimes _{i=1}^n R\left[ \left(\vect{1}M^j\right)_i, \left(\vect{b}\hat{M}^{j}\right)_i   \right] 
\end{equation}
where
\begin{equation}\label{ecR2}
 \begin{array}{ccc}
R\left(0,0\right)&=&1 \\
R\left(0,1\right)&=&H \\
R\left(1,0\right)&=&1 \\
R\left(1,1\right)&=&H T^\dagger 
 \end{array}
\end{equation}
and the subindex indicates the qubit in which each operator is acting.

So this transformed the first operator of the stabilizer into a product of $Z$ and $1$ for each single qubit using $O(n)$ quantum gates. The fourth step is to transform this product into a $Z$ on the first qubit and identities on every other qubit. This is easily done via successive control-not conjugations, each one with control on each of the qubits with $Z$, except for the first one, and target on the first qubit.

So far we have conjugated the chosen operator via $U_1$ defined as:
 \begin{equation}
\begin{split}
 U_1=\left\lbrace\prod_{i=2}^n \left(\mathrm{C-Not}\left( i, 1 \right)\right)^{\left(1-\delta_{\left(1M^j\right)_i,0}\delta_{\left(\vec{b}\hat{M}^{j}\right)_i,0} \right)}\right\rbrace\\
\times\left\lbrace\bigotimes _{i=1}^n R\left[ \left(1M^j\right)_i,\left( \vec{b}\hat{M}^{j}\right)_i    \right]\right\rbrace 
\end{split}
\end{equation}
transforming the first generator of the stabilizer group of the basis chosen into $Z_1$, using $O(n)$ quantum gates. However, all the other operators on the stabilizer group are changed due to the conjugation performed by $U_1$. The next step is to see how they are changed. This is easily done in the circuit formalism, constructing the circuit $U_1 P U_1^{\dagger}$ and permuting the C--Not gates and the rotations with the Pauli operators in $P$ can be done easily, and requires $O(n)$ classical operations for each of the $n-1$ remaining operators in $G_{\vec{b}}$.

It should be noticed, as it will be used later on, that the order of the generators is conserved. That is, the first generator of a stabilizer group corresponding to $j=0$ will be transformed into the first generator of another stabilizer group, and so on.

We have found a way to transform the generators of $G_{\vec{b}}$ into a the generators of a group $\tilde G_{\vec{b}}$ that has $Z_1$ as one of the stabilizers. So every other operator in $\tilde G_{\vec{b}}$ can have either a $Z$ or a $1$ on the first qubit. The remaining canonical generators (i.e., those with $j=1,...,n-1$) will transform into operators having the identity on the first qubit. This way, the generator set $G_{\vec{b}}$ has been transformed into $Z_1$ and the generators of a stabilizer group of $n-1$ qubits, in $O(n^2)$ classical operations and $O(n)$ quantum gates. Now we have to repeat the procedure from the second step and on for the remaining $n-1$ qubits to obtain the circuit for changing from the basis chosen into the computational basis. It's easy to see that the whole circuit will require $O(n^3)$ classical operations for its construction, and $O(n^2)$ quantum gates.

\subsection{An enlightening example}

We now illustrate the ideas of the method through an example. 
The zeroth step of the method is the choice of a primitive polynomial for the field $GF\left(2^n\right)$. In this example we will consider 3 qubits, i.e. $n=3$, and the one chosen for this example is $P\left(x\right)=1+x+x^3$. This polynomial gives the following companion matrix:

\begin{equation}
 M=\left( \begin{array}{ccc}
           0&1&0 \\
	   0&0&1 \\
	   1&1&0  \end{array}\right)
\end{equation}

And each basis of the MUB will be characterized by a $\vec{b}$, as shown on equation \eqref{ecEstab}. In this example we will consider the basis with $\vec{b}=\left(1,0,1 \right)$. It is straightforward to find the corresponding stabilizer group generators:

\begin{eqnarray}
 P_{\vec{b},0}&=&Y\otimes 1 \otimes Z\\
 P_{\vec{b},1}&=&1\otimes Y \otimes Z\\
 P_{\vec{b},2}&=&Z\otimes Z \otimes Y
\end{eqnarray}

\begin{figure}[ht]
\begin{equation*}
\Qcircuit @C=1em @R=.7em {
     &\qw & \targ      & \gate{H}   & \gate{T} & \gate{Y} & \gate{T^{\dagger}} & \gate{H} & \targ    &\qw&& &&\gate{Z}&\qw&        \\ 
     &\qw & \qw        & \qw        & \qw      & \gate{1} & \qw                & \qw      & \qw      &\qw&&=&&\gate{1}&\qw&        \\
     &\qw & \ctrl{-2}  & \qw        & \qw      & \gate{Z} & \qw                & \qw      & \ctrl{-2}&\qw&& &&\gate{1}&\qw&
}
\end{equation*}
\caption{Example of a first step of a change of basis circuit.}
\label{figEjCambioDeBase1}
\end{figure}
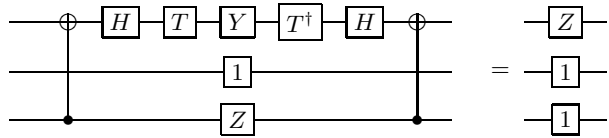

Once the generators for the stabilizer group are known we can follow the previously described steps. 
Take the first operator. Since it has a $Y$ on the first qubit, conjugate it via $T^{\dagger}H$. Then apply a C-Not with control in the third qubit and target in the first to transform this first operator into $Z\otimes 1 \otimes 1$ as shown on figure \ref{figEjCambioDeBase1}.

Next we have to see how the other generators of the stabilizer group transform under this circuit. It is not difficult to see that the transformed stabilizer group $G'$ is:

\begin{equation}
 G'=\left\lbrace Z\otimes 1 \otimes 1 , 1 \otimes Y \otimes Z, 1\otimes Z \otimes X \right\rbrace
\end{equation}

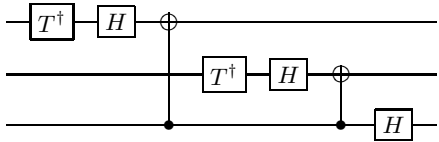
\begin{figure}[ht]
\begin{equation*}
\Qcircuit @C=1em @R=.7em {
     & \gate{T^{\dagger}} & \gate{H} & \targ    &\qw                  &\qw     &\qw      &\qw     &\qw&  \\ 
     & \qw                & \qw      & \qw      &\gate{T^{\dagger}}   &\gate{H}&\targ    &\qw     &\qw& \\
     & \qw                & \qw      & \ctrl{-2}&\qw                  &\qw     &\ctrl{-1}&\gate{H}&\qw&
}
\end{equation*}
\caption{Example of a change of basis circuit.}
\label{figEjCambioDeBase2}
\end{figure}

We have to repeat the above procedure for the the last two generators. This will not modify the first generator since it only acts on the first qubit. This defines the recursive procedure that will, at last, generate the change of basis circuit, that will be the right side of the circuits. In figure \ref{figEjCambioDeBase2} the full result is shown.

The iterative procedure described to generate the circuit is applied $n$ times.
Each iterative step incorporates $O\left(n\right)$ quantum gates, which means the full circuit will use $O\left(n^2\right)$ single and two-qubit quantum gates.
On the other hand, the classical overhead for obtaining a description of the circuit to apply requires $O\left(n^3\right)$ classical processing, giving rise to the dominant term in the efficiency of our method.

\end{document}